\RequirePackage{lineno}
\documentclass[aps,prl,superscriptaddress,twocolumn,showpacs,preprintnumbers,amsmath,amssymb]{revtex4}
\usepackage{epsfig}

\def\met{\mbox{${\hbox{$E$\kern-0.6em\lower-.1ex\hbox{/}}}_T$}} 
\def\mex{\mbox{${\hbox{$E$\kern-0.6em\lower-.1ex\hbox{/}}}_x$}} 
\def\mey{\mbox{${\hbox{$E$\kern-0.6em\lower-.1ex\hbox{/}}}_y$}} 
\def\mexy{\mbox{${\hbox{$E$\kern-0.6em\lower-.1ex\hbox{/}}}_{x,y}$}} 
 
\def\D0{D\O}                            
\begin{document}
\lefthyphenmin=2
\righthyphenmin=3

%
%

\title{A search for high-mass resonances decaying to dimuons at CDF  }

\affiliation{Institute of Physics, Academia Sinica, Taipei, Taiwan 11529, Republic of China} 
\affiliation{Argonne National Laboratory, Argonne, Illinois 60439} 
\affiliation{University of Athens, 157 71 Athens, Greece} 
\affiliation{Institut de Fisica d'Altes Energies, Universitat Autonoma de Barcelona, E-08193, Bellaterra (Barcelona), Spain} 
\affiliation{Baylor University, Waco, Texas  76798} 
\affiliation{Istituto Nazionale di Fisica Nucleare Bologna, $^v$University of Bologna, I-40127 Bologna, Italy} 
\affiliation{Brandeis University, Waltham, Massachusetts 02254} 
\affiliation{University of California, Davis, Davis, California  95616} 
\affiliation{University of California, Los Angeles, Los Angeles, California  90024} 
\affiliation{University of California, San Diego, La Jolla, California  92093} 
\affiliation{University of California, Santa Barbara, Santa Barbara, California 93106} 
\affiliation{Instituto de Fisica de Cantabria, CSIC-University of Cantabria, 39005 Santander, Spain} 
\affiliation{Carnegie Mellon University, Pittsburgh, PA  15213} 
\affiliation{Enrico Fermi Institute, University of Chicago, Chicago, Illinois 60637} 
\affiliation{Comenius University, 842 48 Bratislava, Slovakia; Institute of Experimental Physics, 040 01 Kosice, Slovakia} 
\affiliation{Joint Institute for Nuclear Research, RU-141980 Dubna, Russia} 
\affiliation{Duke University, Durham, North Carolina  27708} 
\affiliation{Fermi National Accelerator Laboratory, Batavia, Illinois 60510} 
\affiliation{University of Florida, Gainesville, Florida  32611} 
\affiliation{Laboratori Nazionali di Frascati, Istituto Nazionale di Fisica Nucleare, I-00044 Frascati, Italy} 
\affiliation{University of Geneva, CH-1211 Geneva 4, Switzerland} 
\affiliation{Glasgow University, Glasgow G12 8QQ, United Kingdom} 
\affiliation{Harvard University, Cambridge, Massachusetts 02138} 
\affiliation{Division of High Energy Physics, Department of Physics, University of Helsinki and Helsinki Institute of Physics, FIN-00014, Helsinki, Finland} 
\affiliation{University of Illinois, Urbana, Illinois 61801} 
\affiliation{The Johns Hopkins University, Baltimore, Maryland 21218} 
\affiliation{Institut f\"{u}r Experimentelle Kernphysik, Universit\"{a}t Karlsruhe, 76128 Karlsruhe, Germany} 
\affiliation{Center for High Energy Physics: Kyungpook National University, Daegu 702-701, Korea; Seoul National University, Seoul 151-742, Korea; Sungkyunkwan University, Suwon 440-746, Korea; Korea Institute of Science and Technology Information, Daejeon, 305-806, Korea; Chonnam National University, Gwangju, 500-757, Korea} 
\affiliation{Ernest Orlando Lawrence Berkeley National Laboratory, Berkeley, California 94720} 
\affiliation{University of Liverpool, Liverpool L69 7ZE, United Kingdom} 
\affiliation{University College London, London WC1E 6BT, United Kingdom} 
\affiliation{Centro de Investigaciones Energeticas Medioambientales y Tecnologicas, E-28040 Madrid, Spain} 
\affiliation{Massachusetts Institute of Technology, Cambridge, Massachusetts  02139} 
\affiliation{Institute of Particle Physics: McGill University, Montr\'{e}al, Qu\'{e}bec, Canada H3A~2T8; Simon Fraser University, Burnaby, British Columbia, Canada V5A~1S6; University of Toronto, Toronto, Ontario, Canada M5S~1A7; and TRIUMF, Vancouver, British Columbia, Canada V6T~2A3} 
\affiliation{University of Michigan, Ann Arbor, Michigan 48109} 
\affiliation{Michigan State University, East Lansing, Michigan  48824}
\affiliation{Institution for Theoretical and Experimental Physics, ITEP, Moscow 117259, Russia} 
\affiliation{University of New Mexico, Albuquerque, New Mexico 87131} 
\affiliation{Northwestern University, Evanston, Illinois  60208} 
\affiliation{The Ohio State University, Columbus, Ohio  43210} 
\affiliation{Okayama University, Okayama 700-8530, Japan} 
\affiliation{Osaka City University, Osaka 588, Japan} 
\affiliation{University of Oxford, Oxford OX1 3RH, United Kingdom} 
\affiliation{Istituto Nazionale di Fisica Nucleare, Sezione di Padova-Trento, $^w$University of Padova, I-35131 Padova, Italy} 
\affiliation{LPNHE, Universite Pierre et Marie Curie/IN2P3-CNRS, UMR7585, Paris, F-75252 France} 
\affiliation{University of Pennsylvania, Philadelphia, Pennsylvania 19104}
\affiliation{Istituto Nazionale di Fisica Nucleare Pisa, $^x$University of Pisa, $^y$University of Siena and $^z$Scuola Normale Superiore, I-56127 Pisa, Italy} 
\affiliation{University of Pittsburgh, Pittsburgh, Pennsylvania 15260} 
\affiliation{Purdue University, West Lafayette, Indiana 47907} 
\affiliation{University of Rochester, Rochester, New York 14627} 
\affiliation{The Rockefeller University, New York, New York 10021} 
\affiliation{Istituto Nazionale di Fisica Nucleare, Sezione di Roma 1, $^{aa}$Sapienza Universit\`{a} di Roma, I-00185 Roma, Italy} 

\affiliation{Rutgers University, Piscataway, New Jersey 08855} 
\affiliation{Texas A\&M University, College Station, Texas 77843} 
\affiliation{Istituto Nazionale di Fisica Nucleare Trieste/Udine, $^{bb}$University of Trieste/Udine, Italy} 
\affiliation{University of Tsukuba, Tsukuba, Ibaraki 305, Japan} 
\affiliation{Tufts University, Medford, Massachusetts 02155} 
\affiliation{Waseda University, Tokyo 169, Japan} 
\affiliation{Wayne State University, Detroit, Michigan  48201} 
\affiliation{University of Wisconsin, Madison, Wisconsin 53706} 
\affiliation{Yale University, New Haven, Connecticut 06520} 
\author{T.~Aaltonen}
\affiliation{Division of High Energy Physics, Department of Physics, University of Helsinki and Helsinki Institute of Physics, FIN-00014, Helsinki, Finland}
\author{J.~Adelman}
\affiliation{Enrico Fermi Institute, University of Chicago, Chicago, Illinois 60637}
\author{T.~Akimoto}
\affiliation{University of Tsukuba, Tsukuba, Ibaraki 305, Japan}
\author{B.~\'{A}lvarez~Gonz\'{a}lez}
\affiliation{Instituto de Fisica de Cantabria, CSIC-University of Cantabria, 39005 Santander, Spain}
\author{S.~Amerio$^w$}
\affiliation{Istituto Nazionale di Fisica Nucleare, Sezione di Padova-Trento, $^w$University of Padova, I-35131 Padova, Italy} 

\author{D.~Amidei}
\affiliation{University of Michigan, Ann Arbor, Michigan 48109}
\author{A.~Anastassov}
\affiliation{Northwestern University, Evanston, Illinois  60208}
\author{A.~Annovi}
\affiliation{Laboratori Nazionali di Frascati, Istituto Nazionale di Fisica Nucleare, I-00044 Frascati, Italy}
\author{J.~Antos}
\affiliation{Comenius University, 842 48 Bratislava, Slovakia; Institute of Experimental Physics, 040 01 Kosice, Slovakia}
\author{G.~Apollinari}
\affiliation{Fermi National Accelerator Laboratory, Batavia, Illinois 60510}
% ...

\collaboration{CDF Collaboration\footnote{With visitors from $^a$University of Massachusetts Amherst, Amherst, Massachusetts 01003,
$^b$Universiteit Antwerpen, B-2610 Antwerp, Belgium, 
$^c$University of Bristol, Bristol BS8 1TL, United Kingdom,
$^d$Chinese Academy of Sciences, Beijing 100864, China, 
$^e$Istituto Nazionale di Fisica Nucleare, Sezione di Cagliari, 09042 Monserrato (Cagliari), Italy,
$^f$University of California Irvine, Irvine, CA  92697, 
$^g$University of California Santa Cruz, Santa Cruz, CA  95064, 
$^h$Cornell University, Ithaca, NY  14853, 
$^i$University of Cyprus, Nicosia CY-1678, Cyprus, 
$^j$University College Dublin, Dublin 4, Ireland,
$^k$Royal Society of Edinburgh/Scottish Executive Support Research Fellow,
$^l$University of Edinburgh, Edinburgh EH9 3JZ, United Kingdom, 
$^m$Universidad Iberoamericana, Mexico D.F., Mexico,
$^n$Queen Mary, University of London, London, E1 4NS, England,
$^o$University of Manchester, Manchester M13 9PL, England, 
$^p$Nagasaki Institute of Applied Science, Nagasaki, Japan, 
$^q$University of Notre Dame, Notre Dame, IN 46556,
$^r$University de Oviedo, E-33007 Oviedo, Spain, 
$^s$Texas Tech University, Lubbock, TX  79409, 
$^t$IFIC(CSIC-Universitat de Valencia), 46071 Valencia, Spain,
$^u$University of Virginia, Charlottesville, VA  22904,
$^{cc}$On leave from J.~Stefan Institute, Ljubljana, Slovenia, 
}}
\noaffiliation

%
%
\begin{abstract}
 We present a search for  high-mass neutral resonances using dimuon 
 data corresponding to an integrated luminosity of 2.3  fb$^{-1}$
 collected in 
 {\mbox{$p\bar p$}}\ collisions at {\mbox{$\sqrt{s}$ =\ 1.96\ TeV}} by
 the CDF II detector at the Fermilab Tevatron.  No significant excess above the standard model expectation
 is observed in the dimuon invariant-mass spectrum. We set 95\% confidence level upper limits on 
 $\sigma \cdot BR (p \bar{p} \rightarrow X \rightarrow \mu \bar{\mu} )$, where $X$ is a boson with spin 0, 1, or
 2. Using these cross section limits, we  determine lower mass limits on  
 sneutrinos in $R$-parity-violating supersymmetric models, $Z'$ bosons, and 
 Kaluza-Klein gravitons in the Randall-Sundrum model.  
\end{abstract}



\maketitle

\vfill\eject
 Neutral resonances decaying to muons have historically been a source of
major discoveries. They also occur in a variety of theoretical models which attempt
to unify the standard model (SM) forces or explain the large gap between the SM
and  gravitational energy scales. The gauge group $SU(3)_C \times SU(2)_L \times U(1)_Y$ of the SM can be 
 embedded in larger gauge groups such as $SU(5)$, $SO(10)$, and $E_6$, to achieve  
 unification in a grand unified theory (GUT)~\cite{rosner, rizzo, langacker, paul2}.  
 In many schemes of GUT symmetry-breaking, $U(1)$ gauge groups
 survive to relatively low energies~\cite{rizzo}, leading to the prediction of  neutral
 gauge vector ($Z^\prime$) bosons. Such $Z^\prime$ bosons typically couple with 
 electroweak strength to SM fermions, and can be observed at hadron colliders as narrow, spin-1, dimuon resonances
 from $q \bar{q} \rightarrow Z^\prime \rightarrow \mu \bar{\mu}$. Many other
 models, such as the $SU(2)_L \times SU(2)_R \times U(1)_{B-L}$ gauge group of the left-right model~\cite{mohapatra}, 
 and the ``little Higgs'' models~\cite{nima, heather}, also predict heavy neutral gauge bosons. 

Additional spatial dimensions are a possible explanation for the gap between the electroweak symmetry-breaking scale 
 and the gravitational energy scale $M_{\mathrm{Planck}}$~\cite{add,rs}. The Randall-Sundrum (RS) 
 model~\cite{rs} 
 predicts excited Kaluza-Klein modes of the graviton, which appear as spin-2 resonances $G^*$ 
 in the process $q \bar{q} \rightarrow G^* \rightarrow \mu \bar{\mu}$. These modes have a narrow intrinsic width when 
  $k/M_{\mathrm {Planck}} <0.1$,  where $k^2$ is the spacetime curvature in the extra dimension.
 In superstring theories with $\cal{O}$(1) couplings, $k/M_{\mathrm {Planck}} \approx 0.01$~\cite{kovermpl}.

 Spin-0 resonances such as the sneutrino $\tilde{\nu}$ in the process $q \bar{q} \rightarrow \tilde{\nu}, \tilde{\bar{\nu}} \rightarrow \mu \bar{\mu}$ are predicted    by 
 supersymmetric theories with $R$-parity violation~\cite{majhi}.  
 Scalar Higgs bosons can be produced as resonances and decay to
 dimuons.  

 The most sensitive direct searches for high-mass boson resonances, which have previously been performed at the Tevatron,   
 have set 95\% confidence level (C.L.) lower limits on the masses $M_{Z^\prime}$, $M_{G^*}$, and $M_{\tilde{\nu}}$ of $Z^\prime$ bosons, 
  RS gravitons, and sneutrinos, respectively. The previous dimuon publication from CDF II, based on $\approx 200$ pb$^{-1}$ of integrated 
 luminosity~\cite{previousCDF}, set mass limits
 that vary from 170 GeV to 885 GeV~\cite{coords} depending on the boson spin and couplings to the SM fermions. 
 Other dilepton and diphoton decay channels have also been
 explored at the Tevatron~\cite{otherChannelsCDF,d0graviton}. Using an order of magnitude more data, we present  
 in this Letter the most sensitive direct search to date for $Z^\prime$, $G^*$, and $\tilde{\nu}$ bosons at high mass. 
     
\begin{figure}[htpb]
\begin{center}
\epsfig {file=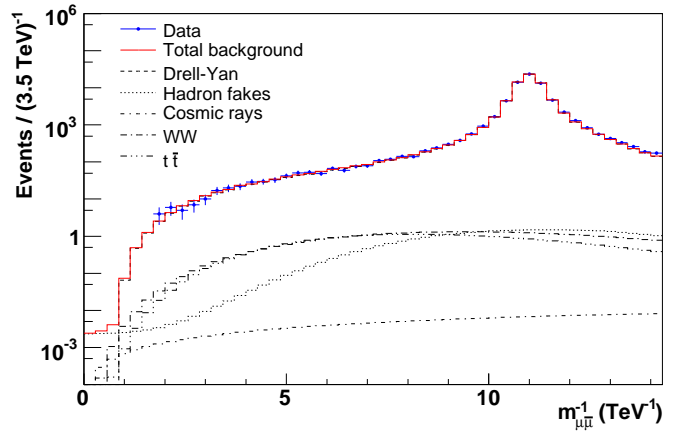,height=6.3cm}
\caption{The distribution of $m_{\mu \bar{\mu}}^{-1}$ (TeV$^{-1}$) for the observed data (points),
 the individual backgrounds (dotted or dashed histograms) and the summed background (solid histogram). The $Z$ boson peak is prominently seen. The inverse mass 
 distribution has the useful feature 
 that the detector resolution is constant ($\approx 0.17$~TeV$^{-1}$) over the range shown in the plot.}
\label{zprimemm_invmass_fit_data}
\end{center}
\end{figure}

 This analysis  uses  2.3~fb$^{-1}$ of data  from  
  $p \bar{p} $ collisions at  $\sqrt{s} = 1.96$~TeV in the CDF II detector~\cite{wzxsection,prd}.
  CDF II is a magnetic spectrometer surrounded by calorimeters
 and muon detectors. We use the 
  central drift chamber
  (COT)~\cite{cot},  the central 
 calorimeter~\cite{run1detector}, and the muon 
 detectors~\cite{muondetectors}  
 for identification and measurement of muons with 
 $| \eta | < 1$~\cite{coords}. 
 The online selection requires a COT track with 
  $p_T > 18 $~GeV \cite{coords}, and 
 matching muon detector hits. 
 We select  a pair of oppositely-charged muons, each 
 with  a COT track with $p_T > 30$~GeV passing quality requirements, and 
 a minimum-ionization signal in the calorimeter. Cosmic rays are rejected
  using  COT
  hit timing~\cite{cotcosmic}. 
 The dimuon signal sample consists of 68150 events in the control region    
   $70 < m_{\mu \bar{\mu}} < 100$~GeV, where the $p \bar{p} \rightarrow Z \rightarrow \mu \bar{\mu}$ process dominates, 
  and 3804 events in the search region $m_{\mu \bar{\mu}} > 100$~GeV. 

  The  alignment of the COT  is performed using a pure sample of high-momentum 
 cosmic-ray muons, in order to obtain the best possible dimuon mass resolution.  Each muon's complete trajectory  is
 fitted to a single helix~\cite{cotcosmic}. The fits are used 
 to determine  the  relative locations of the sense wires, including
 gravitational and electrostatic displacements, with a statistical accuracy of a few microns~\cite{prd}. 
 We constrain remaining misalignments,
 which cause a bias in the track curvature, by comparing  $\left\langle E/p \right\rangle$~\cite{coords} for electrons
 and positrons.  The tracker momentum scale and resolution is measured  
  by template-fitting the $Z \rightarrow \mu \bar{\mu}$  mass peak, and calibrating to
 the  world average values~\cite{lepwmass} of the $Z$ boson mass and width. 

\begin{figure}[htpb]
\begin{center}
 \hspace*{-3mm}
\epsfig {file=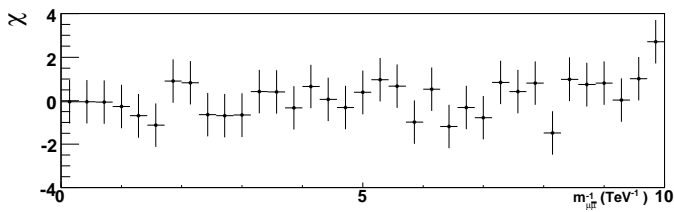,height=2.96cm}
\caption{The difference between the distributions of $m_{\mu \bar{\mu}}^{-1}$ (TeV$^{-1}$) for the observed data and
 the summed background, divided by the expected statistical uncertainty in each bin. All vertical error bars have
 unit size.}
\vspace*{-6mm}
\label{chi_zprimemm_invmass_fit}
\end{center}
\end{figure}

 For a resonance with electroweak coupling and
  mass above 200 GeV, the observed width of the $m_{\mu \bar{\mu}}$ distribution is dominated by the
 track curvature resolution, resulting in an approximately constant resolution of  
  $\delta m_{\mu \bar{\mu}}^{-1} \approx 0.17$~TeV$^{-1}$. 
   Our search strategy is to construct templates of the observable $m_{\mu \bar{\mu}}^{-1}$ distribution for
 a range of 
 boson Breit-Wigner pole masses, add the background distributions to the templates, and compare the templates to 
 the $m_{\mu \bar{\mu}}^{-1}$ distribution
 from the data in the search region 
  $m_{\mu \bar{\mu}} > 100$ GeV. The simulated templates (including backgrounds) are
 normalized to the data in the 70~GeV~$< m_{\mu \bar{\mu}} < 100$~GeV region, thus cancelling several sources
 of systematic uncertainty. 

We determine the most likely number of signal events ($N_S$),  
and the corresponding confidence intervals~\cite{feldmanCousins},  from the binned Poisson likelihood~\cite{prd} for the observed data to be produced by a
sum of signal and background templates. 
 The use of the constant-resolution variable $m_{\mu \bar{\mu}}^{-1}$ simplifies the optimization of the
 template binning and the scan over the boson pole masses. 
 
 Signal and SM Drell-Yan background distributions are evaluated 
 using a specialized  Monte Carlo (MC)
 simulation~\cite{prd}  
 of boson production and decay, and of the detector response to the 
 leptons and hadrons. 
 The  kinematics of boson production and decay are obtained
 from the {\sc pythia}~\cite{pythia} event generator
 using the CTEQ6M~\cite{cteq6} set of parton distribution functions. 
 QED radiation is simulated~\cite{prd} based on the {\sc wgrad}
  program~\cite{wgrad}. 
  The  MC performs a detailed hit-level simulation
 of the lepton tracks. COT
 hits are generated  according to their   
 resolution ($\approx 150 ~ \mu$m) and measured efficiencies, and 
 a helix fit is performed (as it is in data) to simulate the reconstructed track. 
 We apply a mass-dependent next-to-next-to-leading
 order (NNLO) multiplicative correction ($K$-factor)~\cite{carena} to the SM Drell-Yan background. 

 The SM production processes for $W^+W^-$~\cite{wwnlo} 
 and $t \bar{t}$~\cite{ttbarnlo} 
 have small contributions, and are evaluated using  their NLO cross sections, {\sc pythia}, and a detector
 simulation based on {\sc geant}~\cite{geant}.   
 Misidentification backgrounds result from cosmic rays, QCD jets, and
 $\pi/K$ decays-in-flight (DIF). We evaluate the cosmic-ray background using a large sample of
 cosmic rays identified with the COT-timing-based algorithm~\cite{cotcosmic}, and using the direction-of-flight information
 provided by this algorithm. 
 The $m_{\mu \bar{\mu}}^{-1}$ 
 shape of  misidentified jets is evaluated from a large 
 sample of inclusive jet events. Decays-in-flight within the COT active volume generate a kink
 along the helical trajectory, resulting in a mismeasurement of the track curvature. For large reconstructed
  momenta, the measured DIF curvature distribution is approximately uniform and leads to
 a flat  $m_{\mu \bar{\mu}}^{-1}$ spectrum. 
 Most DIF tracks are rejected using 
 their abnormal COT-hit pattern and large fit $\chi^2$. 
The jet and DIF backgrounds are normalized
 using the mass distribution of same-charge dimuon events. 

 Figure~\ref{zprimemm_invmass_fit_data} shows the $m_{\mu \bar{\mu}}^{-1}$ distributions of the observed data and the expected
 backgrounds, which are in good agreement (as shown in  Fig.~\ref{chi_zprimemm_invmass_fit}). 
 A resonance whose observed width is dominated by detector resolution would appear as a peak spanning
 approximately three bins. The likelihood-based fitter finds no significant excess. We use background-only ensembles of 
 simulated events, each with the statistics of the data sample, to evaluate the
 probability of statistical fluctuations anywhere in the search region generating a discrepancy at least as significant as the largest discrepancy found in the
 data. We find this probability (``p-value'') to be 6.6\% and we conclude that the observed data are statistically consistent with the 
 SM expectation. The dielectron $m_{ee}$ spectrum from  2.5~fb$^{-1}$ of CDF II data~\cite{dielectronPRL}
 shows that the largest discrepancy with the expected background occurs
 at $m_{ee} \sim 240$~GeV. Figure~\ref{chi_zprimemm_invmass_fit} shows that the dimuon data are consistent
 with the expectation near this mass to better than 1$\sigma$ in statistical precision. The sensitivity of the
 dielectron analysis for a spin-1 resonance at this mass is $\approx 20$\%  better than the dimuon analysis reported here. 

\begin{figure}[t]
\begin{center}
\epsfysize = 5.1cm
\epsffile {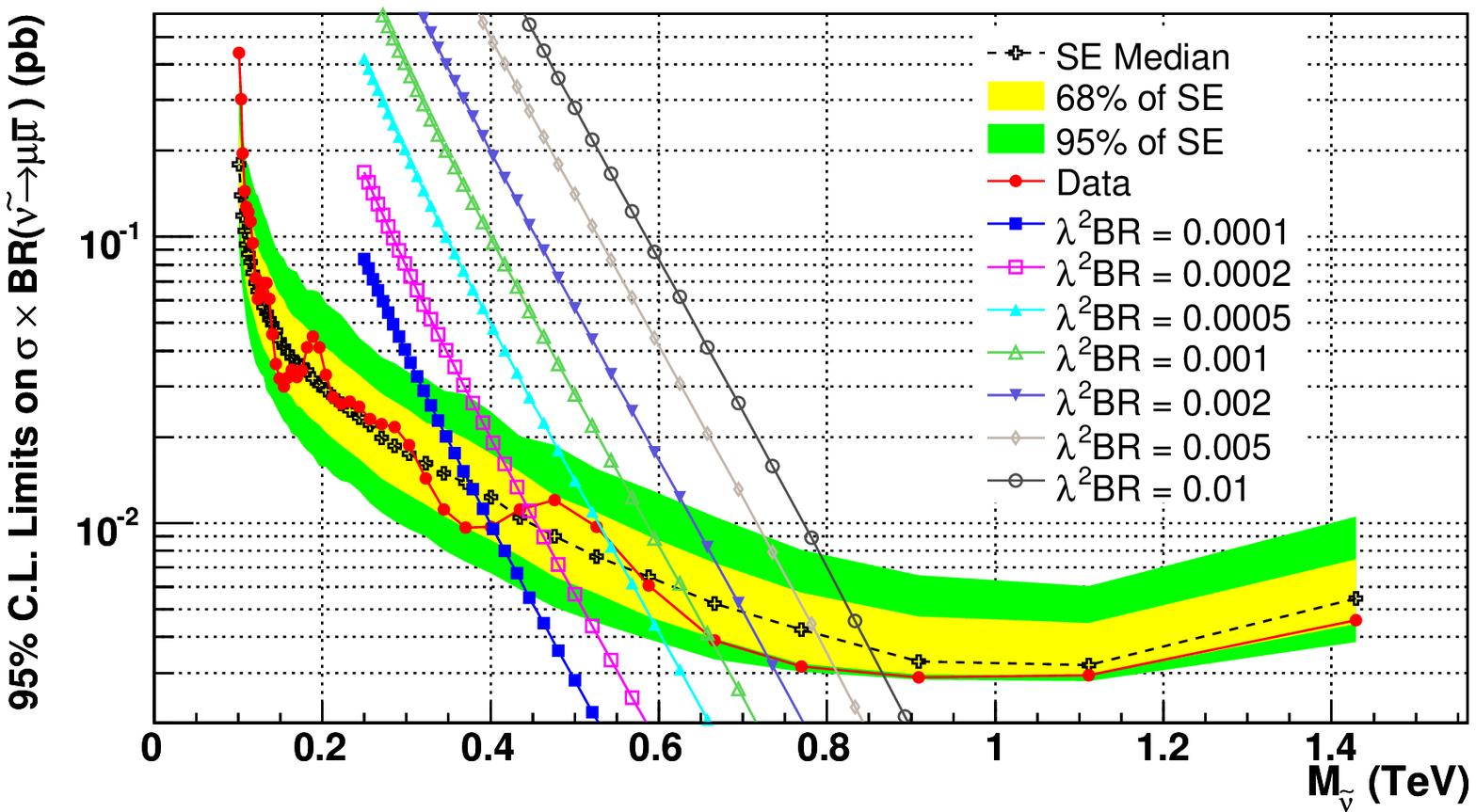}
\epsfysize = 5.1cm
\epsffile {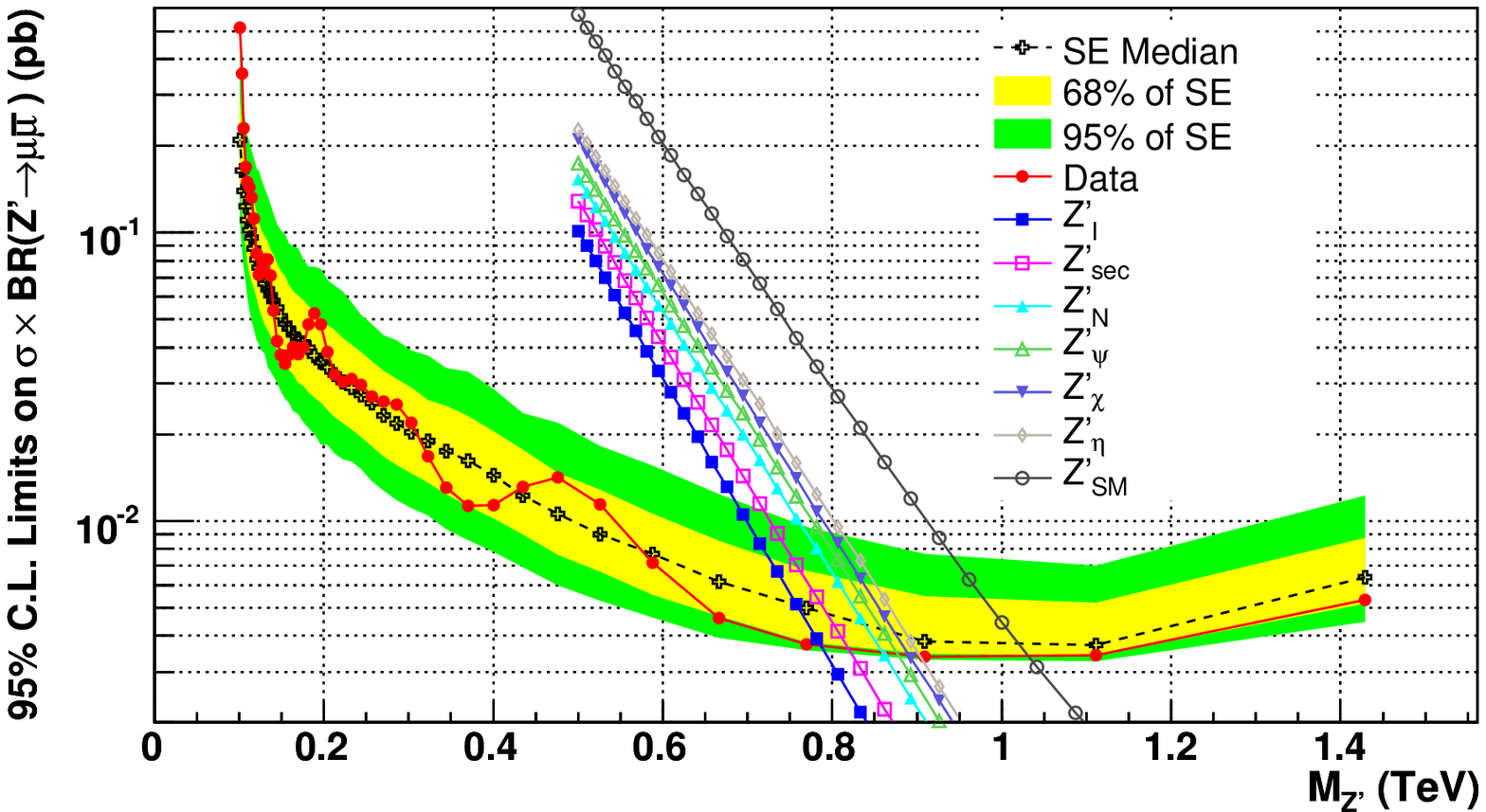}
\epsfysize = 5.1cm
\epsffile {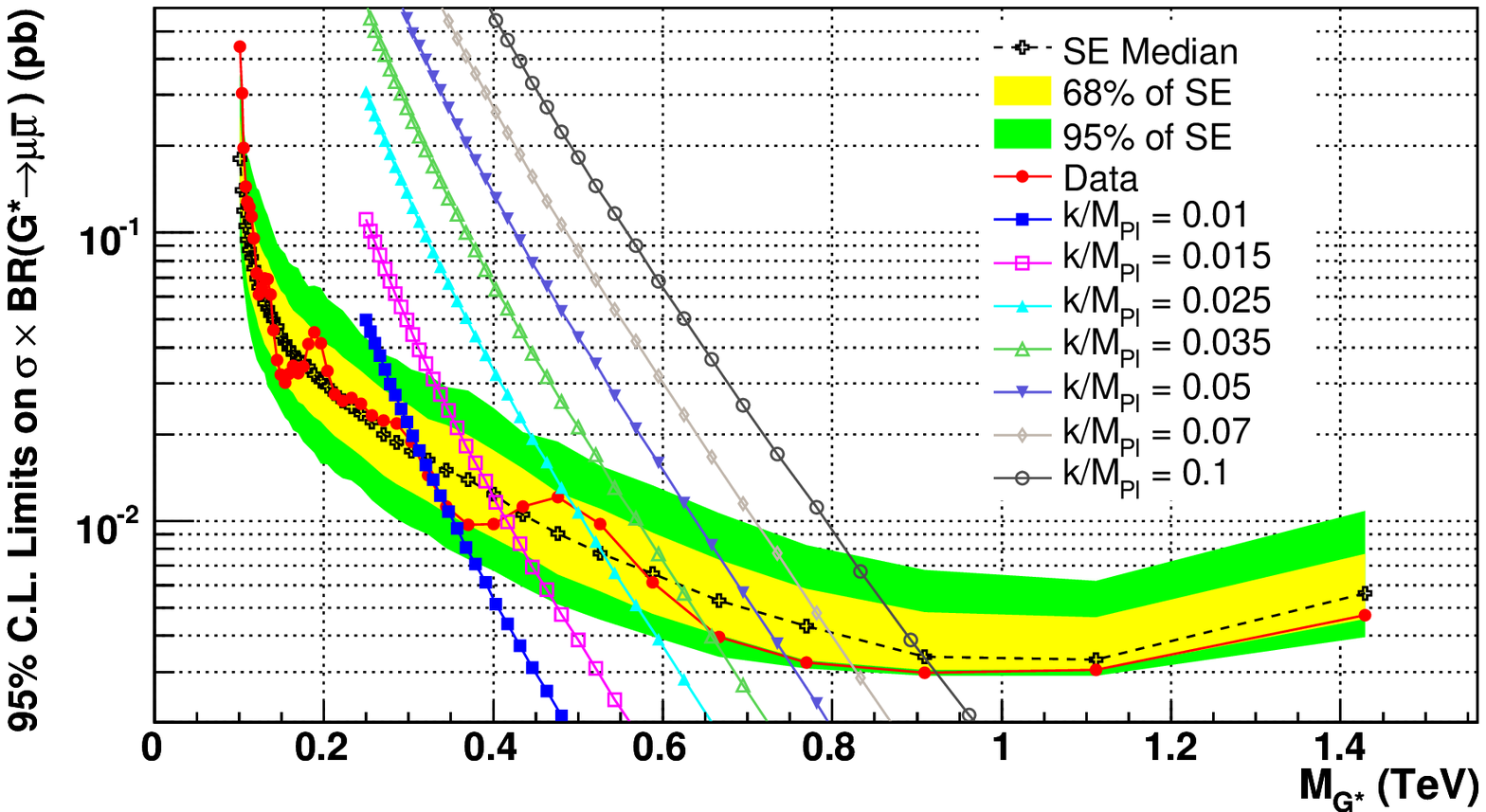}
\caption{ The 95\% C.L. upper limits on 
$\sigma \cdot BR (\tilde{\nu}, \tilde{\bar{\nu}} \rightarrow \mu \bar{\mu})$ vs $M_{\tilde{\nu}}$
(top), 
 $\sigma \cdot BR (Z^\prime \rightarrow \mu \bar{\mu})$ versus $M_{Z'}$
(middle), and  
 $\sigma \cdot BR (G^* \rightarrow \mu \bar{\mu})$ versus $M_{G^*}$
(bottom). Also shown are the theoretical cross
 sections for various model parameter values~\cite{rs,majhi,zprimeModels}.
The expected limits and ranges of limits, as derived from simulated
 experiments (SE), are shown for comparison. The step size between adjacent templates
 in the signal scan is 0.2 TeV$^{-1}$ in pole mass.}
\vspace*{-9mm}
\label{sigmaLimits}
\end{center}
\end{figure}

The likelihood fitter determines the 95\% C.L. upper limit on the number of signal events, 
 for each value of the resonance pole mass. We convert these limits to limits on 
$\sigma \cdot BR (\tilde{\nu}, \tilde{\bar{\nu}} \rightarrow \mu \bar{\mu})$, 
$\sigma \cdot BR (Z^\prime  \rightarrow \mu \bar{\mu})$, and $\sigma \cdot BR (G^* \rightarrow \mu \bar{\mu})$  
 using the total acceptance
 as a function of pole mass, the NNLO cross section for $Z \rightarrow \mu \mu$ of 251.3 pb~\cite{wzxsection}, 
 and dividing by the observed number of $Z \rightarrow \mu \bar{\mu}$ events. The
 acceptance is verified with the detailed {\sc geant}-based simulation, and comparisons
 to data distributions. The muon identification efficiency is verified using a pure data sample 
 of $Z$ bosons triggered by one identified muon. The total acceptance, including kinematic and fiducial acceptance
 and dimuon identification, increases from $\approx 13$\% ($\approx 20$\%) for
 a pole mass of 90 GeV to $\approx 40$\% ($\approx 45$\%) 
 for a $Z^\prime$ (graviton) pole mass of 1 TeV, and decreases for higher pole masses due to the kinematic limit of the
 parton collisions.  
  The 95\% C.L. upper limits on 
 $\sigma \cdot BR (\tilde{\nu}, \tilde{\bar{\nu}} \rightarrow \mu \bar{\mu})$, 
 $\sigma \cdot BR (Z^\prime \rightarrow \mu \bar{\mu})$, and $\sigma \cdot BR (G^* \rightarrow \mu \bar{\mu})$ 
 are shown
 in Fig.~\ref{sigmaLimits}. The dominant mass-dependent systematic uncertainties 
 arise from parton distribution functions (16\%), the NNLO 
 $K$-factor (9\%)~\cite{carena}, QED 
 radiative corrections (3\%)~\cite{baur}, and acceptance (3\%), all quoted at 1 TeV.
  These uncertainties  are incorporated as functions of $ m_{\mu \bar{\mu}} $ and increase monotonically
 beyond 100 GeV. Uncertainties on the momentum scale and resolution, and  on the
  non-Drell-Yan 
 background predictions, have a negligible effect.

\begin{table}[htpb]
\begin{center}
\begin{tabular}{cc|cc|cc}
\hline
\hline
 $Z^\prime$  &  $Z^\prime$    & RS graviton  &  graviton & $\tilde{\nu}$ & $\tilde{\nu}$\\
 model  &  mass limit   & $k/M_{\mathrm{Planck}}$  &  mass limit & $\lambda^2 \cdot BR$ & mass limit\\
\hline
  $Z^\prime_{I} $     &  789    & 0.01  & 293 & 0.0001 & 397\\
  $Z^\prime_{\mathrm sec}$    &  821   &  0.015 & 409 & 0.0002 & 441\\
 $Z^\prime_{N}     $  &  861   &  0.025 & 493 & 0.0005 &  541\\
 $Z^\prime_{\psi}  $  &  878   &  0.035 & 651 & 0.001 &  662\\
 $Z^\prime_{\chi}  $  &  892   &  0.05  & 746 & 0.002 &  731\\
 $Z^\prime_{\eta}  $  &  904   &  0.07  & 824 & 0.005 &  810\\
 $Z^\prime_{SM}    $  & 1030   &  0.1   & 921 & 0.01 &  866\\
\hline
\hline
\end{tabular}
\caption{95\% C.L. lower limits on $Z^\prime$, graviton, and sneutrino masses (in GeV) for various model 
parameters~\cite{rs,majhi,zprimeModels}. For the $R$-parity-violating sneutrino model, $\lambda$ is the
 $d \bar{d} \tilde{\nu}$ coupling and $BR$ denotes the $\tilde{\nu}, \tilde{\bar{\nu}} \rightarrow \mu \bar{\mu} $ branching ratio. }
\label{tab:massLimits}
\end{center}
\end{table}

Our signal templates have been generated with a resonance pole width $\Gamma =  2.8\% \times M$, based on the 
 SM $Z$ boson width. Thus our signal scan probes an 
 observed width of $\approx [17\% (M/{\rm TeV}) \oplus 2.8\%] ~ M$.
In a model where the observed width increases by a factor $x$, the cross section limits would increase by about a factor of $\sqrt{x}$.

 We use {\sc pythia} to 
  compute the cross sections for production of $Z^\prime$ bosons predicted by  $E_6$ models~\cite{zprimeModels} or
 having the same couplings to SM fermions as the $Z$ boson, and of $G^*$ bosons for various $k / M_{\mathrm{Planck}}$ values. We apply the
 NNLO $K$-factor to these LO cross sections. The NLO $\tilde{\nu}$ production cross sections are obtained
 from~\cite{majhi}. 
 We derive the boson mass limits shown in
 Table~\ref{tab:massLimits}. 

In conclusion, we have presented a direct search for  high-mass neutral resonances with spin-0, 1, and 2, using 
 an integrated luminosity of 
 2.3 fb$^{-1}$  collected by the CDF II detector. Our dimuon
 invariant mass spectrum is consistent with the SM expectation. We set the world's tightest constraints on $Z^\prime$ bosons in various models, on Kaluza-Klein
 graviton modes in the RS model, and on sneutrinos in $R$-parity violating supersymmetric models.
 At 95\% C.L., 
 we exclude $100 < M_{Z^\prime} < 982$ GeV for a $Z^\prime_{\eta}$ boson of the $E_6$ model,
  $100 < M_{G^*} < 921$ GeV for $k/M_{\mathrm{Planck}} = 0.1$, 
 and  $100 < M_{\tilde{\nu}} < 810$ GeV for $\lambda^2 \cdot BR (\tilde{\nu}, \tilde{\bar{\nu}} \rightarrow \mu \bar{\mu})=0.01$,
 where  $\lambda$ is the
 $d \bar{d} \tilde{\nu}$ coupling and $BR$ denotes the $\tilde{\nu}, \tilde{\bar{\nu}} \rightarrow \mu \bar{\mu} $ branching ratio. 

We thank the Fermilab staff and the technical staffs of the participating institutions for their vital contributions. 
 This work was supported by the U.S. Department of Energy and National Science Foundation; the Italian Istituto 
 Nazionale di Fisica Nucleare; the Ministry of Education, Culture, Sports, Science and Technology of Japan; 
 the Natural Sciences and Engineering Research Council of Canada; the National Science Council of the Republic 
 of China; the Swiss National Science Foundation; the A.P. Sloan Foundation; the Bundesministerium f\"ur Bildung 
 und Forschung, Germany; the Korean Science and Engineering Foundation and the Korean Research Foundation; the
 Science and Technology Facilities Council and the Royal Society, UK; the Institut National de Physique Nucleaire 
 et Physique des Particules/CNRS; the Russian Foundation for Basic Research; the Ministerio de Ciencia e 
 Innovaci\'{o}n, Spain; the Slovak R\&D Agency; and the Academy of Finland.


\begin{references} 
%
\bibitem{rosner} J.~L.~Rosner,  Phys. Rev. D {\bf 35}, 2244 (1987).
%
\bibitem{rizzo} F.~del Aguila, M.~Quiros, and F.~Zwirner, Nucl. Phys. 
{\bf B287}, 419 (1987);
J.~L.~Hewett, and T.~G.~Rizzo, Phys. Rep. {\bf 183}, 193 (1989).
%
\bibitem{langacker} 
J.~Erler {\it et al.}, Phys. Rev. D {\bf 66}, 015002 (2002);
T.~Han {\it et al.}, Phys. Rev. D {\bf 70}, 115006 (2004).
%
\bibitem{paul2} J.~Kang and P.~Langacker, Phys. Rev. D {\bf 71}, 035014 
(2005).
%
\bibitem{mohapatra}
R. N. Mohapatra and J. C. Pati, Phys. Rev. D {\bf 11},
566 (1975); G. Senjanovic and R. N. Mohapatra, Phys. Rev. D {\bf 12},
1502 (1975); R. N. Mohapatra and G. Senjanovic, Phys. Rev. D {\bf 23},
165 (1981).
%
\bibitem{nima} 
N.~Arkani-Hamed {\it et al.}, J. High Energy Phys. 07, 034 (2002).

\bibitem{heather} T.~Han {\it et al.}, Phys. Rev. D {\bf 67}, 095004 (2003).
%
\bibitem{add}
N. Arkani-Hamed, S. Dimopoulos, and G. Dvali, Phys. Lett. B {\bf 429}, 263 (1998). 
%
\bibitem{rs}
L. Randall and R. Sundrum, Phys. Rev. Lett. {\bf 83}, 4690 (1999); 
L. Randall and R. Sundrum, Phys. Rev. Lett. {\bf 83}, 3370 (1999). 
%
\bibitem{kovermpl}
 H.~Davoudiasl, J.~L.~Hewett and T.~G.~Rizzo,
         Phys. Rev. Lett.  {\bf 84}, 2080 (2000). 
%
\bibitem{majhi} 
D. Choudhury, S. Majhi, and V. Ravindran, Nucl. Phys. B {\bf 660}, 343 (2003). We assume the $\tilde {\nu}$ and $\tilde{\bar{\nu}}$ have equal masses and couplings and contribute equally to a potential signal, as in 
  W.~Shao-Ming, H.~Liang, M.~Wen-Gan, Z.~Ren-You and J.~Yi,
  Phys.\ Rev.\  D {\bf 74}, 057902 (2006).
%
\bibitem{previousCDF}
 A.~Abulencia {\it et al.} (CDF Collaboration),
Phys. Rev. Lett. {\bf 95}, 252001 (2005).
%
\bibitem{coords}  We use the convention $\hbar = c = 1$. 
 Pseudorapidity is defined as 
 $\eta = -\ln[\tan(\theta/2)$], where $\theta$ is the polar angle from the beam axis. Track momentum $p$  transverse to
  the beam is denoted $p_T$. $E$ denotes the calorimeter energy. 
%
\bibitem{otherChannelsCDF}
 T.~Aaltonen {\it et al.} (CDF Collaboration), Phys. Rev. Lett. {\bf 99}, 171802 (2007); 
 T.~Aaltonen {\it et al.} (CDF Collaboration), Phys. Rev. Lett. {\bf 99}, 171801 (2007); 
 A.~Abulencia {\it et al.} (CDF Collaboration), Phys. Rev. Lett. {\bf 96}, 211801 (2006); 
 D.~Acosta {\it et al.} (CDF Collaboration), Phys. Rev. Lett. {\bf 95}, 131801 (2005).
%
\bibitem{d0graviton}
 V.~M. Abazov {\it et al.} (D0 Collaboration), 
 Phys. Rev. Lett. {\bf 100}, 091802 (2008). 
%
\bibitem {wzxsection}
 A.~Abulencia {\it et al.} (CDF Collaboration), J. Phys. G {\bf 34}, 2457 (2007); 
D. Acosta {\it et al.} (CDF Collaboration),  Phys. Rev. Lett. {\bf 94}, 091803 (2005). 
%
\bibitem {prd}  T.~Aaltonen {\it et al.} (CDF Collaboration),  Phys. Rev. D {\bf 77}, 112001 (2008).
%
\bibitem{cot}
T. Affolder {\it et al.}, Nucl. Instrum. Methods A {\bf 526}, 249 (2004).
%
\bibitem{run1detector}
 F. Abe {\it et al.} (CDF Collaboration),
 Nucl. Instrum. Methods A {\bf 271}, 387 (1988).
%
\bibitem{muondetectors}
 G. Ascoli  {\it et al.}, Nucl. Instrum. Methods  A {\bf 268}, 33 (1988).
%
\bibitem{cotcosmic}
A.~V.~Kotwal, H.~K.~Gerberich, and C.~Hays,  Nucl. Instrum. Methods A {\bf 506}, 110 (2003).
%
\bibitem{lepwmass} S. Schael {\it et al.} (ALEPH, DELPHI, L3, OPAL and SLD Collaborations), Phys. Rep. {\bf 427},
 257 (2006)  and references therein.
%
\bibitem{feldmanCousins}
G. J. Feldman and R. D. Cousins, Phys. Rev. D {\bf 57}, 3873 (1998).
%
\bibitem{pythia}
T.~Sj$\rm\ddot{o}$strand, Comput. Phys. Commun. {\bf 82}, 74 (1994), version 6.208.
%
\bibitem{cteq6} J. Pumplin {\it et al.}, J. High Energy Phys. {\bf 0207}, 012 (2002).
%
\bibitem {wgrad}  U. Baur, S. Keller, and D. Wackeroth,  Phys. Rev. D  {\bf 59}, 013002 (1998).
%
\bibitem{carena}
M. Carena, A. Daleo, B. A. Dobrescu, and T. M. P. Tait, Phys. Rev. D {\bf 70}, 093009 (2004).
%
\bibitem{wwnlo}
J. M. Campbell and R. K. Ellis, Phys. Rev. D {\bf 60}, 113006 (1999), {\sc mcfm} version 3.4.5.
%
\bibitem{ttbarnlo}
R. Bonciani, S. Catani, M. L. Mangano, and P. Nason, Nucl. Phys. B {\bf 529}, 424 (1998); M. Cacciari
{\it et al.}, J. High Energy Phys. 04, 068 (2004).
%
\bibitem{geant}
R.~Brun and F.~Carminati, CERN Program Library Long Writeup, W5013, 1993
(unpublished), version 3.15.
%
\bibitem{dielectronPRL}
T. Aaltonen {\it et al.}  (CDF Collaboration), Phys. Rev. Lett. 102, 031801 (2009).
%
\bibitem{baur}
  U.~Baur, O.~Brein, W.~Hollik, C.~Schappacher, and D.~Wackeroth,
  Phys.\ Rev.\  D {\bf 65}, 033007 (2002), Fig. 3(b).
%
\bibitem{zprimeModels}
 The couplings we use are detailed in C.~Ciobanu {\it et al.}, FERMILAB-FN-0773-E (2008).
%
\end{references}
\end{document}